\documentstyle[12pt,epsfig]{article}
\newcommand{\beq}{\begin{equation}}
\newcommand{\eeq}{\end{equation}}
\newcommand{\beqn}{\begin{eqnarray}}
\newcommand{\eeqn}{\end{eqnarray}}

\newcommand{\stackm}{\stackrel{\scriptstyle <}{{ }_{\sim}}}
\newcommand{\sbt}{{\tilde b}}

\newcommand{\pl}{P_L}
\newcommand{\pr}{P_R}
 
\begin{document}

\thispagestyle{empty}
\def\pubnum{XXX}
\def\data{November, 1997}
\begin{flushright}
{\parbox{3.5cm}{
UAB-FT-430\\ KA-TP-1-1998

January, 1998

hep-ph/9802329
}}
\end{flushright}
\vspace{3cm}
\begin{center}
\begin{large}
\begin{bf}
YUKAWA-COUPLING CORRECTIONS TO SCALAR QUARK DECAYS\\
\end{bf}
\end{large}
\vspace{1cm}
Jaume GUASCH 
\,,
Joan SOL{\`A}\\
\vspace{0.25cm}
Grup de F{\'\i}sica Te{\`o}rica\\
 
and\\
 
Institut de F{\'\i}sica d'Altes Energies\\
 
\vspace{0.25cm} 
Universitat Aut{\`o}noma de Barcelona\\
08193 Bellaterra (Barcelona), Catalonia, Spain\\
\vspace{1cm}
Wolfgang HOLLIK\\
Institut f{\"u}r Theoretische Physik,\\
 Universit{\"a}t Karlsruhe,
D-76128 Karlsruhe, Germany\\
\end{center}
\vspace{0.3cm}
\hyphenation{super-symme-tric co-lli-ders}
\hyphenation{com-pe-ti-ti-ve e-le-men-ta-ry}
\hyphenation{coun-ter-terms}
\begin{center}
{\bf ABSTRACT}
\end{center}
\begin{quotation}
\noindent
Heavy squark decays into top and charginos or neutralinos could be an
unexpected source of top quarks at hadron colliders.
A detailed treatment of these processes is necessary for a reliable 
calculation of both the top quark production cross-section 
and the standard top quark branching ratio in the MSSM.
Along this line we compute the electroweak corrections  to
the sbottom decays $\tilde{b}_a\rightarrow \chi^-_i\,t $ within the Yukawa
coupling approximation. 
The calculation of these higher order contributions requires
renormalization of both the bottom squark mixing angle and of
$\tan\beta$. This type of corrections gives the leading order 
electroweak effects at low and high $\tan\beta$.

\end{quotation}
  
\newpage

\baselineskip=6.5mm  %(FOR PREPRINT)

A fundamental missing link between the highly successful Standard Model
of the Elementary Particle Physics and the highly developed status of the
experiments is the final elucidation of the nature of the spontaneous
symmetry-breaking mechanism (SSB).
Whether the understanding of the SSB will eventually point towards the
existence of elementary spinless particles (Higgs bosons) or towards 
the finding of dynamically  bound
fermion-antifermion spinless states, is far from clear at present.
Ultimately, experiment must decide. In the meanwhile, the Minimal
Supersymmetric extension of the Standard Model (MSSM) remains 
immaculately consistent with all known high precision experiments
at a level comparable to the SM\,\cite{WdeBoer}. This fact alone,
if we bare in mind the vast amount of high precision data available at present
both from low-energy and high-energy physics, 
should be considered at least as greatly notorious. To this fact we have to
add another prominent feature, to wit: that the MSSM offers
a starting point for a successful Grand Unified framework where a
radiatively stable low-energy Higgs sector can survive.
Putting things together it is  well justified, we believe, to keep alive all
efforts on all fronts trying to discover a supersymmetric 
particle. Undoubtedly, the next Tevatron
run, and of course also the advent of the LHC, should offer us a gold-plated
scenario for finding a hint of SUSY, if it is there at all.    

Sparticles not much heavier than a few hundred $GeV$
could be produced in significant numbers already at the Tevatron. For instance, 
selectron production was advocated in Ref.\cite{Ambrosanio} to explain a
purported non-SM event in the Collider Detector at Fermilab (CDF). 
Subsequently, in Ref.\cite{KaneMrenna} it was argued that half of the 
top quarks at the Tevatron might come from gluino decays into
top and stop, $\tilde{g}\rightarrow t\,\tilde{t}_1$. Similarly, we may
envision the possibility that sbottom squarks are pair produced by the
usual Drell-Yan mechanism and then decay into top quark and charginos:
$\tilde{b}_a\rightarrow\chi^-_i\,t$. Indeed, this would be the leading
two-body decay if gluinos are heavy enough that the strong decay mode
$\tilde{b}_a\rightarrow \tilde{g}\,b$ is kinematically
blocked up\footnote{Squark decays have been discussed at the tree-level
in several places of the literature. See e.g. Ref.\cite{BM} 
for some relatively recent references on the subject.}.
The observed
cross-section is then equal to the Drell-Yan production cross-section
convoluted over the parton distributions times the
squared branching ratio. However, in the framework of the MSSM, we rather
expect a generalization of this formula in the following way (schematically),
\beqn
\sigma_{\rm obs.}&=&\int dq\,d\bar{q}\,\,\sigma (q\,\bar{q}\rightarrow t\,\bar{t})\,
\times |BR (t\rightarrow W^+\,b)|^2\nonumber\\
&+& \int dq\,d\bar{q}\,\,\sigma (q\,\bar{q}\rightarrow 
\tilde{b}_a\,\bar{\tilde{b}}_a)\,
\times |BR (\tilde{b}_a\rightarrow\chi^-_1\,t)|^2\times
|BR (t\rightarrow W^+\,b)|^2+...
\label{eq:productionMSSM}
\eeqn
We assume that gluinos are much heavier than squarks,
so that their contribution to this cross-section
through $q\,\bar{q}\rightarrow 
\tilde{g}\,\tilde{g}$ followed by
 $\tilde{g}\rightarrow t\,\tilde{t}_1$ is negligible.
From eq.(\ref{eq:productionMSSM}) we see that,
if there are alternative (non-SM) sources of top quarks subsequently
decaying into the SM final state, $W^+\,b$, then one cannot rigorously
place any stringent lower bound on
$BR (t\rightarrow W^+\,b)$ in the MSSM from the present FNAL data.
Indeed, we could as well have non-SM top quark decay modes, such as e.g.
$t\rightarrow \tilde{t}_a\,\chi^0_{\alpha}$\,\cite{KaneMrenna} and 
$t\rightarrow H^+\,b$\,\cite{GuaschSola,CGGJS}, that could
serve, pictorially, as a ``sinkhole''  to compensate (at least in part) for 
the unseen source of extra top quarks produced at the Tevatron from sbottom 
pair production (Cf. eq.(\ref{eq:productionMSSM})).
However, one usually assumes that
 $BR (t\rightarrow W^+\,b)\geq 50\%$ in order to guarantee the
(purportedly) standard top quark events observed at the Tevatron. 
Thus, from these considerations it is not excluded
that the non-SM branching ratio of the top quark, 
$BR (t\rightarrow $``new''$)$,
could be as big as the SM one, i.e. $\sim 50\%$.
We stress that at present one cannot exclude eq.(\ref{eq:productionMSSM})
since the observed $t\rightarrow W^+\,b$ final state
involves missing energy, as it is also the case for the decays
comprising supersymmetric particles.
In particular, if $\tan\beta$ is large and there exists a relatively light
chargino with a non-negligible higgsino component, the alternative mechanism
suggested in eq.(\ref{eq:productionMSSM}) could be a rather efficient
non-SM source of top quarks that could compensate for the depletion in
the SM branching ratio. 

While the squark production cross-section has already  received some
attention in the literature at the level of NLO radiative corrections
\,\cite{Beenakker}, an accurate treatment of the
decay mechanisms is also very important to provide a solid basis
for experimental analyses of the top quark production in the MSSM. 
Thus in this paper we consider the computation of the 
leading supersymmetric electroweak (SUSY-EW)
quantum effects on $\tilde{b}_a\rightarrow\chi^-_i\,t$, namely the
ones induced by potentially large Yukawa-couplings from the top and bottom
quarks  (normalized with respect to the $SU(2)_L$ gauge coupling):
\begin{equation}
\lambda_t\equiv {h_t\over g}={m_t\over \sqrt{2}\,M_W\,\sin{\beta}}\;\;\;\;\;,
\;\;\;\;\; \lambda_b\equiv {h_b\over g}={m_b\over \sqrt{2}
\,M_W\,\cos{\beta}}\,,
\label{eq:Yukawas} 
\end{equation}
At large ($\geq 20$) or small ($<1$) $\tan\beta$\,\cite{Hunter} 
these effects could be
competitive with the known QCD corrections\,\cite{DHJ}. Since
in these conditions the full MSSM
quantum effects can be rather large, their calculation is indispensable 
to account for the observed 
top quark production cross-section (\ref{eq:productionMSSM}) in the MSSM
or, alternatively, to better assess how much the determination of the SM
branching ratio $BR(t\rightarrow W^+\,b)$ is affected in the MSSM context 
after plugging in the experimental number on 
the LHS of eq.(\ref{eq:productionMSSM}).

We address the calculation of the one-loop corrections to
the partial width
of $\,\tilde{b}_a\rightarrow\chi^-_i\,t$  within the context 
of the on-shell renormalization framework\,\cite{BSH}.
However, in SUSY extensions of the SM,
there are  additional ingredients
concerning the renormalization program that must be fixed. 
Consider the MSSM interaction Lagrangian  involving the 
top-sbottom-chargino vertex in the
mass-eigenstate basis: 
\beq
  {\cal L}_{t\sbt\chi}=-g\, \sbt^*_a 
\bar{\chi}^+_i (A_{+}^{ai}\pl+\epsilon_i\, 
A_{-}^{ai}\pr) t +{\rm h.c.}\,,
\label{eq:ltl}
\eeq
where $\epsilon_i$ is the sign of the {\it i}th chargino eigenvalue $M_i\,
(i=1,2$ with $|M_1|<|M_2|$) in the real matrix representation, and the coupling
matrices are denoted by\footnote{See Refs.\cite{CGGJS,Guasch1} for
full notation niceties.}
\beq
  A_{+}^{ai}=R_{1a} V_{i1}-\lambda_b R_{2a} V_{i2}\,,\ \ \ \
  A_{-}^{ai}=-R_{1a} \lambda_t U_{i2}\,.
\label{eq:AS}
\eeq
The explicit appearance of the Yukawa couplings (\ref{eq:Yukawas})
in the above Lagrangian requires both the
introduction of top and bottom quark mass counterterms
(in the on-shell scheme) and also a suitable prescription for 
$\tan\beta$ renormalization. 
We denote by $m_{\tilde{b}_a}\,(a=1,2)$, with
$m_{\tilde{b}_1}<m_{\tilde{b}_2}$, the two sbottom mass eigenvalues.
The sbottom mixing angle $\theta_{\tilde{b}}$
is defined by the transformation relating the weak-interaction
($\sbt^\prime_a=\sbt_L, \sbt_R$) and the mass eigenstate
 ($\sbt_a=\sbt_1, \sbt_2$) squark bases:
\beq
\label{eq:defsq}
  \sbt^\prime_a=R_{ab}\, \sbt_b\,\,; \,\,\,\,
  R=\left(\begin{array}{cc}
      \cos\theta_{\tilde{b}}&-\sin\theta_{\tilde{b}} \\
      \sin\theta_{\tilde{b}}&\cos\theta_{\tilde{b}}
    \end{array}\right)\,,
\eeq
where $R$ is the matrix appearing in eq.(\ref{eq:AS}).
From this change of basis the sbottom mass matrix,
\begin{equation}
{\cal M}_{\tilde{b}}^2 =\left(\begin{array}{cc}
M_{\tilde{b}_L}^2+m_b^2+\cos{2\beta}(-{1\over 2}+
{1\over 3}\,s_W^2)\,M_Z^2 
 &  m_b\, (A_b-\mu\tan\beta)\\
m_b\, (A_b-\mu\tan\beta) &
M_{\tilde{b}_R}^2+m_b^2-{1\over 3}\,\cos{2\beta}\,s_W^2\,M_Z^2\,,  
\end{array} \right)\,,
\label{eq:sbottommatrix}
\end{equation}
becomes diagonalized as follows: 
$R^{\dagger}\,{\cal M}_{\tilde{b}}^2\,R=
{\rm diag}\left\{m_{\tilde{b}_2}^2, m_{\tilde{b}_1}^2\right\}\,.
$

Formally, we start the renormalization procedure as usual\,\cite{BSH} i.e. by 
introducing parameter and field renormalization constants:
\begin{eqnarray}
  g&\rightarrow&(1+\frac{\delta g}{g})\,g\,,  \nonumber\\
                m_q&\rightarrow&(1+\frac{\delta m_q}{m_q})\,m_q\,,\nonumber\\
  t&\rightarrow&(1+\frac{1}{2}\delta Z_L^t \pl+\frac{1}{2}
\delta Z_R^t \pr)\,t\,,\nonumber\\
  \chi^+_i&\rightarrow&(1+\frac{1}{2}\delta Z_L^i \pl+\frac{1}{2}
\delta Z_R^i \pr)\,
  \chi^+_i\,,\nonumber\\
  \sbt_a&\rightarrow&(1+\frac{1}{2} \delta Z^a)\,\sbt_a + 
\delta Z^{ab} \sbt_b\ \ \ (b\neq a)\,,
  \nonumber\\
  \lambda_{b,t}&\rightarrow&(1+\frac{\delta
    \lambda_{b,t}}{\lambda_{b,t}})\,\lambda_{b,t}\,, \nonumber\\
  \tan\beta&\rightarrow&(1+\frac{\delta\tan\beta}{\tan\beta})\,
\tan\beta\,,\nonumber\\
  R&\rightarrow&R+\delta R\,.
\label{eq:shifts}
\end{eqnarray}
Notice that the sbottom wave-function renormalization involves the
mixed scalar field renormalization constant ($b\neq a$)
\beq
\delta Z^{ab}
={\Sigma^{ab}(m_{\tilde{b}_b}^2)
\over m_{\tilde{b}_b}^2-m_{\tilde{b}_a}^2}\,,
\label{eq:renzeta}
\eeq
where $\Sigma^{ab}$ is the bare mixed self-energy connecting the physical states
$\tilde{b}_a$ and $\tilde{b}_b$ ($a\neq b$).
The renormalization of the Yukawa couplings (\ref{eq:Yukawas})
involves the mass counterterms and the counterterm to
$\tan\beta$ through the relations
\begin{eqnarray}
  \frac{\delta\lambda_b}{\lambda_b}&=&\frac{\delta m_b}{m_b}-\frac{\delta
  M_W}{M_W}+\sin^2\beta\,\frac{\delta \tan\beta}{\tan\beta}\,,\nonumber\\
  \frac{\delta\lambda_t}{\lambda_t}&=&\frac{\delta m_t}{m_t}-\frac{\delta
  M_W}{M_W}-\cos^2\beta\,\frac{\delta \tan\beta}{\tan\beta}\,.
\label{eq:deltes}
\end{eqnarray}
Furthermore, as the last row of eq.(\ref{eq:shifts}) shows, we 
have introduced
a counterterm $\delta R$ for the rotation matrix $R$ that diagonalizes the sbottom
mass matrix.
This counterterm can be replaced (see below) by a counterterm 
$\delta\theta_{\tilde{b}}$ for the sbottom mixing angle.
Indeed, we shall proceed this way as
we wish to take $\theta_{\tilde{b}}$ as an input in our 
renormalization program.

Substituting the relations (\ref{eq:shifts}) into the
bare Lagrangian (\ref{eq:ltl}) we find: 
\beq
  {\cal L}_{\rm one-loop}=-g\, \sbt^*_a \bar{\chi}^+_i  \left[
    \left(A_{+}^{ai}+\delta C^{ai}_{+}\right)\pl + \epsilon_i\left(A_{-}^{ai}
+\delta C^{ai}_{-}\right)\pr\right] t+{\rm h.c.}\,,
\eeq
with
\beqn
\delta C^{ai}_{+}&=&A_{+}^{ai}\left(
\frac{\delta g}{g}+\frac{1}{2}\delta Z^{a} +\frac{1}{2}
      \delta Z_R^i +\frac{1}{2}\delta Z_L^t\right)
+\delta A_{+}^{ai}+\delta Z^{ba} A_{+}^{bi}\,,\nonumber\\
\delta C^{ai}_{-}&=&A_{-}^{ai}\left(
\frac{\delta g}{g}+\frac{1}{2}\delta Z^{a} +\frac{1}{2}
      \delta Z_L^i +\frac{1}{2}\delta Z_R^t\right)
+\delta A_{-}^{ai}+\delta Z^{ba} A_{-}^{bi}\,.
\label{eq:dcounter}
\eeqn
The full structure of the four on-shell renormalized decay amplitudes 
for $\tilde{b}_a\rightarrow\chi^-_i\, t\, (a=1,2; i=1,2)$
follows from the previous Lagrangian after including the contributions 
from the (LH and RH)
one-loop vertex form factors $F_{L,R}^{ai}$:
\beq
i\,T(\sbt_a\rightarrow t\, \chi^-_i )    = i\,g\,\bar
  u_t\,\left[\epsilon_i\left(A_{-}^{ai}+\Lambda_{L}^{ai}\right)\pl
+\left(A_{+}^{ai}+\Lambda_R^{ai}\right)\pr\right]
  \,v_i
\eeq
where
\beq
\Lambda_{L}^{ai}=\delta C_{-}^{ai}+\epsilon_i\,F_{L}^{ai}\,,\ \ \ \ \
\Lambda_{R}^{ai}=\delta C_{+}^{ai}+F_{R}^{ai}\,.
\eeq
Let now $\Gamma_0^{ai}$ be the
tree-level partial width of the decay $\tilde{b}_a\rightarrow\chi^-_i\,t$
obtained from the Lagrangian (\ref{eq:ltl}):
\beq
\Gamma_0^{ai} =\frac{g^2}{16\,\pi\,m_{\tilde{b}_a}^3}\,\lambda (a,i,t)\,
\left\{\left[(A_{+}^{ai})^2+(A_{-}^{ai})^2\right]\,(m_{\tilde{b}_a}^2-M_i^2-m_t^2)
-4\,A_{+}^{ai}\,A_{-}^{ai}\,m_t\,\epsilon_i\,M_i\right\}\,,
\label{eq:tree}
\eeq 
with $\lambda (a, i, t)\equiv\lambda(m_{\tilde{b}_a}^2, M_i^2, m_t^2)$ the usual
K{\"a}llen function for the given arguments.
The quantum correction to $\Gamma_0^{ai}$ can be
described in terms of the quantities
$\delta^{ai}=(\Gamma^{ai} - \Gamma_0^{ai})/\Gamma_0^{ai}$, where $\Gamma^{ai}$
is the corresponding one-loop corrected width.
From the previous formulae, $\delta^{ai}$ can be worked out
as follows:
\beqn
  \delta^{ai}=\frac{2 \left(m_{\tilde{b}_a}^2-M_i^2-m_t^2\right)
\left( A_{-}^{ai}\,
\Lambda_L^{ai}+A_{+}^{ai}\,\Lambda_R^{ai} \right)-4\,m_t\,\epsilon_i\,M_i\,(
    A_{-}^{ai}\Lambda_R^{ai}+A_{+}^{ai}\,\Lambda_L^{ai})}
  {\left[(A_{+}^{ai})^2+(A_{-}^{ai})^2\right]
\left(m_{\tilde{b}_a}^2-M_i^2-m_t^2\right)
-4\,A_{+}^{ai}\,A_{-}^{ai}\,m_t\,\epsilon_i\,M_i}\,.
\label{eq:correction}
\eeqn
%Now, although this expression is formally correct, we still 
For the further procedure, we
have to fix the
renormalization conditions that determine the counterterms $\delta\tan\beta$
and $\delta\theta_{\tilde{b}}$ which enter the explicit structure of
$\delta A_{\pm}^{ai}$ in eq.(\ref{eq:dcounter}).
Following Ref.\cite{CGGJS} we
use the decay process $H^{+}\rightarrow\tau^{+}\nu_{\tau}$ 
to define the parameter $\tan\beta$ through
\beq
\Gamma(H^{+}\rightarrow\tau^{+}\nu_{\tau})=
{g^2 m_{\tau}^2\,M_H\over 32\,\pi\, M_W^2}\,\tan^2\beta= 
{G_F m_{\tau}^2\,M_{H}\over 4\pi\sqrt{2}}\,\tan^2\beta\, 
(1-\Delta r^{MSSM})\,.
\label{eq:tbetainput}
\eeq
This determines the counterterm $\delta\tan\beta$\,\cite{CGGJS} as follows:
\beq
{\delta\tan\beta\over \tan\beta}
=\frac{1}{2}\left(
\frac{\delta M_W^2}{M_W^2}-\frac{\delta g^2}{g^2}\right)
-\frac{1}{2}\delta Z_{H^\pm}
+\cot\beta\, \delta Z_{HW}+ 
\Delta_{\tau}\,,
\label{eq:deltabeta}
\eeq 
where $\Delta_{\tau}$ stands for the full set of MSSM one-loop corrections
to the partial width of $H^{+}\rightarrow\tau^+\,\nu_{\tau}$.   
%As for $\delta\theta_{\tilde{b}}$, we pin it down upon requiring that the
For fixing $\delta\theta_{\tilde{b}}$, we require that the
renormalized sbottom mixing angle
(that we use as an input data, Cf. eq.(\ref{eq:inputb}) below) does not
%get shifted by 
feel a shift from 
the mixed sbottom bare self-energies $\Sigma^{ab}$ between the
physical states $\tilde{b}_a$ and $\tilde{b}_b$ ($a\neq b$).
This is similar to the prescription adopted in Refs.\cite{DHJ,EBM},
though it is not identical.  
In our formalism, the 3-point Green functions explicitly incorporate 
the mixed field renormalization
constants $\delta Z^{ab}$ ($a\neq b$) and are therefore renormalized also
in the $\theta_{\tilde{b}}$ parameter. The 
UV-divergent parts of the 3-point functions are cancelled against
$\delta\theta_{\tilde{b}}$ by defining the latter as follows:
\beq
\delta\theta_{\tilde{b}}=\frac12\,(\delta Z^{12}-\delta Z^{21})
=\frac12\,{\Sigma^{12}(m_{\tilde{b}_2}^2)+\Sigma^{12}(m_{\tilde{b}_1}^2)
\over m_{\tilde{b}_2}^2-m_{\tilde{b}_1}^2}\,.
\label{eq:rentheta}
\eeq
Of course another equivalent choice could just be
$\delta\theta_{\tilde{b}}=\delta Z^{12}$ (or $-\delta Z^{21}$), but 
%ours, 
eq.(\ref{eq:rentheta}), is more symmetrical; the numerical 
differences among the finite
parts of the two choices are negligible.

%While the 
The analytical formulation developed thus far is well
suited to tackle the general problem of the SUSY-EW
corrections to squark decays. %, in the following 
Since the dominant part is from the Yukawa sector we wish to pursue
our calculation in the following  within the Yukawa coupling approximation.
This means that we are going to compute the leading 
electroweak effects of ${\cal O}(\lambda^2_t)$ 
{\it and} ${\cal O}(\lambda^2_b)$ that emerge 
for large values of the Yukawa couplings (\ref{eq:Yukawas}) when the remaining
gauge contributions -- of ${\cal O}(g^2)$-- are subdominant.
In practice we shall only explore the large $\tan\beta$
 regime, typically $\tan\beta\geq 20$;
the possibility  $\tan\beta<1$ is not so appealing from the theoretical
point of view.  Thus within our approximation we will include the correction 
$\Delta_{\tau}$ in leading 
order ${\cal O}(\lambda_{\tau}^2)$ of the $\tau$ Yukawa-coupling, $\lambda_{\tau}$.
Notice furthermore that for $\lambda_b>>1$ the
tree-level decay rate, eq.\,(\ref{eq:tree}), is maximized. Therefore,
the large $\tan\beta$ range is expected to be the most relevant
one for the decay under consideration. 

%In our approach, we assume that charginos $\chi^{\pm}_i$ and 
%neutralinos $\chi^{0}_\alpha$ are mainly
%higgsinos by setting the $SU(2)_L$ gaugino mass parameter $M>>|\mu|, M_W$ in the chargino mass matrix\,\cite{CGGJS}.
%It is only in this case that the Yukawa-coupling approximation makes
%sense. 
%Thus, since $m_{\tilde{t}_1}>80-90\,GeV$, in this approach the decay
%$t\rightarrow\tilde{t}_a\,\chi^0_{\alpha}$ is kinematically
%forbidden\footnote{See, however, Ref.\cite{GHS2} for a more general
%  discussion.}.
In our approach, we set the $SU(2)_L$ gaugino mass parameter $M>>|\mu|, M_W$ in
the chargino mass matrix\,\cite{CGGJS}, and therefore the chargino $\chi^\pm_1$
is mainly higgsino, whereas the chargino $\chi^\pm_2$ is mainly gaugino and does
not contribute to our decays.
It is only in this case that the Yukawa-coupling approximation makes
sense. 
Thus, since $m_{\tilde{t}_1}>80-90\,GeV$, in this approach the decay into stop
and neutralino
$t\rightarrow\tilde{t}_a\,\chi^0_{\alpha}$ is kinematically
forbidden\footnote{See, however, Ref.\cite{GHS2} for a more general
  discussion.}.
Therefore, in this approximation the relevant counterterms
$\delta A_{\pm}^{ai}$ in eq.(\ref{eq:dcounter}) boil down to 
\beqn
\delta A_{+}^{a2}&=&-\delta R_{2a}  
\lambda_b -R_{1a} \delta \lambda_b\nonumber\\
\delta A_{-}^{a2}&=&-\delta R_{1a}\, 
\lambda_t -R_{1a}\, \delta \lambda_t\,,
\eeqn
with $\delta \lambda_b$ and $\delta \lambda_t$ as in eq.(\ref{eq:deltes}).
The one-loop Feynman diagrams contributing to $\tilde{b}_a\rightarrow\chi^{-}_i\,t$
in our Yukawa-coupling approximation
are depicted in Fig.\,1. Specifically, in Fig.\,1a we show the wave-function
renormalization graphs from which the field renormalization constants
$\delta Z^a$,  $\delta Z^i$ and $\delta Z^t_{L,R}$
are straightforwardly computed; and in
Fig.1b we display the contributions going to the vertex form factors $F^{ai}_{L,H}$.
We refrain from writing out cumbersome formulas in this letter\,\cite{GHS2}.
Nevertheless we remark that the calculation is well-defined
in the Yukawa coupling approximation. In this respect, we have checked the
effective  cancellation
of divergences among the bare diagrams and the various counterterms. 
From the practical point of view 
we emphasize that the definition of $\tan\beta$ that we have used to fix
$\delta\tan\beta$ (Cf. eq.(\ref{eq:tbetainput})) is quite convenient
as the Tevatron (and the LHC) collaborations have started to develop
all the necessary
techniques for $\tau$-tagging and $H^{\pm}$ identification\,\cite{CDF}.

Let us now pass on to the numerical analysis. It is summarized in Figs.\,2-5.
The set of independent parameters that we use in the sbottom sector
% is the following
consists of the masses and the mixing angle
\beq
(m_{\tilde{b}_1}, m_{\tilde{b}_2}, \theta_{\tilde{b}})\,,
\label{eq:inputb}
\eeq
whereas for the stop sector we just have in addition 
\beq
(m_{\tilde{t}_1}, \theta_{\tilde{t}})
\label{eq:inputt}
\eeq
since by $SU(2)_L$ gauge invariance %already determines 
the value of the 
%heaviest 
other
stop mass $m_{\tilde{t}_2}$ is already determined.
Similarly, the sbottom and stop trilinear terms $A_b$ and $A_t$ are fixed by
the previous parameters as follows:
\beq
A_{b}=\mu\,\tan\beta+
{m_{\tilde{b}_2}^2-m_{\tilde{b}_1}^2\over 2\,m_b}\,\sin{2\,\theta_{\tilde{b}}}\,;
\ \ \ \
A_{t}=\mu\,\cot\beta+
{m_{\tilde{t}_2}^2-m_{\tilde{t}_1}^2\over 2\,m_t}\,\sin{2\,\theta_{\tilde{t}}}\,.
\label{eq:Abt}
\eeq
We impose the approximate (necessary) condition
\beq
A_q^2<3\,(m_{\tilde{t}}^2+m_{\tilde{b}}^2+M_H^2+\mu^2)\,,
\label{eq:necessary}
\eeq
where $m_{\tilde{q}}$
is of the order of the average squark masses
for $\tilde{q}=\tilde{t},\tilde{b}$, to avoid colour-breaking minima 
in the MSSM Higgs potential\,\cite{Frere}. 

For the numerical analysis, we assume that the
sbottom masses are not too heavy ($m_{\tilde{b}_1}\leq 350\,GeV$) and that 
the charged Higgs mass, $M_H$,
is such that the decay $t\rightarrow H^+\,b$ is available. In this way the
quantum corrections that we shall evaluate could have an impact already for
the Tevatron physics of top quark production. However, in Ref.\cite{GHS2}   
we include the corresponding study of the radiative corrections to sbottom
decays also in the case of very heavy (``obese'') sbottom 
and charged Higgs bosons ($m_{\tilde{q}}\geq 400\,GeV$, $M_H>m_t$) which should
be relevant for LHC.
Of course, $\tan\beta$ and the SUSY Higgs mixing
parameter $\mu$ are also additional independent inputs for our calculation.
In the relevant large $\tan\beta$ segment under consideration,
namely 
\beq
20\stackm\tan\beta\stackm 40\,,
\label{eq:tansegment}
\eeq
the bottom quark Yukawa coupling $\lambda_b$ 
is comparable to the top quark Yukawa coupling, $\lambda_t$. 
Even though the extreme interval $40<\tan\beta<60$ can be tolerated
by perturbation theory, we shall confine ourselves to the moderate range
(\ref{eq:tansegment}). This is 
necessary to preserve the condition (\ref{eq:necessary}) for the
typical set of sparticle masses used in our analysis. In particular, notice
that with our chosen set of independent inputs in
eqs.(\ref{eq:inputb})-(\ref{eq:inputt})
the $A$-parameters are not free --they are fixed by eq.(\ref{eq:Abt}). 
We point out that the colour stability requirement (\ref{eq:necessary})
could be satisfied independently of $\tan\beta$ if the $A$-parameters would be
chosen directly as a part of the set of inputs and then taken
sufficiently small. Nevertheless this possibility is not so
convenient in our analysis where the sparticle masses are the natural inputs
that we wish to control in order to make sure that
sparticles can be produced and decay at the Tevatron as explained
in connection to eq.(\ref{eq:productionMSSM}).

In Figs.\,2a and 2b we evaluate the branching ratio
\begin{eqnarray}
BR_0(\tilde{b}_a\rightarrow \chi^-_1\,t)&=&
{\Gamma_0(\tilde{b}_a\rightarrow \chi^-_1\,t)\over \Sigma_i \Gamma_i }
\nonumber\\
\Sigma_i \Gamma_i&=& \Sigma_\alpha\Gamma_0(\tilde{b}_a\rightarrow \chi^0_\alpha\,b)+
\Gamma_0(\tilde{b}_a\rightarrow \chi^-_1\,t)\nonumber\\ 
&+&
\Gamma_0(\tilde{b}_a\rightarrow H^-\,\tilde{t}_1) 
+\Sigma_i\,\Gamma_0(\tilde{b}_a\rightarrow \Phi^0_i\,\tilde{b}_b) \,,
\label{eq:BR}
\end{eqnarray}
(where $\Phi_i^0=h^0,H^0,A^0$)
as a function of $\tan\beta$ and $m_{\tilde{b}_1}$, respectively, in order to
%see what are 
illustrate
the relevant intervals for these parameters where to
compute radiative corrections to our decay. In these figures we have also
indicated the ranges of $\tan\beta$ and $m_{\tilde{b}_1}$ excluded by the
condition (\ref{eq:necessary}) for the given values of the other parameters.
The corresponding corrections $\delta^{ai}$~(\ref{eq:correction}) are shown in 
Figs.\,3a and 3b as a function of the lightest stop and sbottom masses,
respectively.
The allowed range for the sbottom and stop mixing angles is conditioned by
the upper bound on the trilinear couplings and is obtained
from eqs.(\ref{eq:Abt}) and (\ref{eq:necessary}). In the case of
$\theta_{\tilde{b}}$, it is rather narrow, so that the physical sbottom
masses basically coincide with
the LH and RH electroweak eigenstates. In the physical
$\theta_{\tilde{b}}$ range, the variation
of the correction (\ref{eq:correction}) is shown in Fig.\,4a.
On the other hand the permitted range for the stop mixing
angle, $\theta_{\tilde{t}}$, is much larger and we have plotted the corrections
within the allowed region in Fig.\,4b. Notice that
the sign of the quantum effects changes
within the domain of variation of $\theta_{\tilde{t}}$.
Finally, we display the evolution of the SUSY-EW
effects as a function of $\tan\beta$ (Fig.\,5a) and of $\mu$ (Fig.\,5b)
within the region of compatibility with the constraint (\ref{eq:necessary}).
 We remark that for 
$|\mu|>120\,GeV$ and $\tan\beta>20$ the corrections can be above $20\%$
near the phase space limit of the 
lightest sbottom decay. However, this effect
has nothing to do with the phase space exhaustion, which is described by
the kinematic function $\lambda (a, i, t)$ on the RHS of the tree-level expression
(\ref{eq:tree}), but rather with the presence of the dynamical factor
in brackets on that equation which also goes to the denominator
of $\delta$ in eq.(\ref{eq:correction}). That factor is fixed by the 
structure of the interaction Lagrangian
of the sbottom decay into charginos and top; and, for the parameters
in Fig.\,5, it turns out to vanish near (actually past) the phase space
limit in the case of the lightest sbottom ($\tilde{b}_1$) decay.
However, this is not so for the heaviest sbottom ($\tilde{b}_2$) as
it is patent in the same figure.

A few more words are in order to explain the origin of the leading
electroweak effects. One could expect that they come from 
the well-known large $\tan\beta$ enhancement stemming
from the chargino-stop corrections to the bottom mass\footnote{See e.g.
\cite{CGGJS} and references therein.} (Cf. Fig.\,1a). 
Nonetheless this is only partially true, for in the present
case the remaining contributions in Fig.\,1
can be sizeable enough.
To be more precise, in the region of the
parameter space that we have dwelled upon
the bottom mass contribution is seen to be
dominant only for the lightest sbottom decay
and for the lowest values of $\tan\beta$ in the
range (\ref{eq:tansegment}). This is indeed the case in Fig.\,4b where
$\tan\beta=20$ and therefore the bottom mass effect modulates the 
electroweak correction in this process and 
$\delta$ becomes essentially an odd function of the stop mixing angle. 
On the other hand, from Fig.\,5a it is obvious that the 
(approximate) linear behaviour on $\tan\beta$ expected from bottom 
mass renormalization becomes
completely distorted by the rest of the contributions,
especially in the high $\tan\beta$ end.
In short, the final electroweak correction cannot be simply ascribed 
to a single renormalization source but to the full Yukawa-coupling combined yield.

In general the SUSY-EW corrections to 
$\Gamma(\tilde{b}_a\rightarrow \chi^-_i\,t)$ are smaller than 
the QCD corrections\,\cite{DHJ}. 
The reason why the electroweak corrections are smaller
is in part due to
the condition (\ref{eq:necessary}) restricting our analysis within the
$\tan\beta$ interval (\ref{eq:tansegment}). From  Figs.\,4 and~5a it is
clear that outside this interval the SUSY-EW contributions could be much higher
and with the same or opposite
sign as the QCD effects depending on the choice of the sign of the 
mixing angles\footnote{In the conditions of Ref.\cite{DHJ} 
the evolution of the QCD corrections with $\tan\beta$ would be very
slow because the assumed sbottom mixing angle is zero. 
In the present case, however, $\theta_{\tilde{b}}\neq 0$ and the QCD 
corrections evolve significantly with $\tan\beta$ but become
saturated\,\cite{GHS2}.}.
Moreover, since we have focused our analysis to sbottom masses
accessible to Tevatron, again the theoretical bound (\ref{eq:necessary})
severely restricts the maximum value of the trilinear couplings and this
prevents the electroweak corrections from being larger. 
This cannot be cured by assuming larger values of $M_H$ and/or of $\mu$
due to our assumption that $t\rightarrow H^+\,b$ is operative and because
$\mu$ directly controls the value of the (higgsino-like) chargino final
state in our decay, so that basically we have  $|\mu|<m_{\tilde{b}_a}-M_H$.
The restriction cannot be circumvented either if we
assume larger values of $m_{\tilde{t}_a}$, for it has been shown
that too heavy stops are incompatible with the CLEO data on
 $b\rightarrow s\,\gamma$
both at low and high $\tan\beta$\,\cite{Ng,CGGJS2}. 
We point out that the MSSM analysis of $b\rightarrow s\,\gamma$ 
also motivated the sign choice 
$A\,\mu<0$ in our numerical calculation\,\cite{CGGJS2}.
Admittedly, the situation with radiative $B$-decays is still under 
study and there are many sources of %error 
uncertainties
that deserve
further experimental consideration. Still, we have used this information
to focus on a limited domain of the MSSM parameter space. 

Finally,  we mention that the treatment of the electroweak corrections to
the closely related decay $\tilde{t}_a\rightarrow\chi^0_{\alpha}\,t$ is similar
to the one presented here, but for lack of space 
it will be presented elsewhere\,\cite{GHS2}. It suffices to say that
the branching ratio and the radiative corrections to that decay
can be sizeable only for small values
of $\tan\beta$, so that this process becomes optimized in a region of
parameter space complementary to the one under consideration.

In summary, the MSSM corrections to squark decays into charginos and
neutralinos can be significant and therefore must be included in any reliable
analysis of top quark physics at the Tevatron within the MSSM. The
main corrections stem from 
the strongly interacting sector of the theory
(i.e. the one involving gluons and gluinos), but also
non-negligible effects may appear from the electroweak sector 
(characterized by chargino-neutralino exchange) at large (or very small)
values of $\tan\beta$. 
Failure of including these corrections in future studies
of top quark physics at the Tevatron, both in the production and decay
mechanisms, might seriously hamper the possibility of
discovering clear-cut traces of SUSY physics from the identification of 
large non-SM quantum corrections in these processes. 
As already stated, we have mainly concentrated on the impact of
these quantum signatures in the physics of
the Tevatron, but important effects are also expected\,\cite{GHS2}  for 
experiments aiming at the production
and decay of ``obese'' squarks  
at the LHC. The latter type of squarks could be free of some of the restrictions that
have been considered for the present calculation.

{\bf Acknowledgements}:

\noindent
Financial support of J.S. by the Spanish Ministerio de Educaci{\'o}n
y Ciencia and of W.H by the DAAD through the Programa de Acciones Integradas
(Acci{\'o}n No. HA1996-0030) is gratefully acknowledged.
The work of JG has been financed by a 
grant of the Comissionat per a Universitats i Recerca, Generalitat de 
Catalunya. J.S. and J.G. have also been partially supported by CICYT 
under project No. AEN93-0474.

%%%%%%%%%%%%%%%%%%%%%%%%%%%%%%%%%%%%%%%%%%%%%%%%%%%%%%%%%%%%%%%%%%%

\vspace{0.75cm}
\begin{center}
\begin{Large}
{\bf Figure Captions}
\end{Large}
\end{center}
\begin{itemize}
\item{\bf Fig.1} SUSY-EW Feynman diagrams, at one-loop order,
correcting the partial width of $\tilde{b}_a\rightarrow\chi_1^-\,t$:
{\bf (a)} Self-energy diagrams; 
{\bf (b)} Vertex contributions. There are other possible diagrams, but
they do not contribute in the Yukawa-coupling approximation.

\item{\bf Fig.2} {\bf (a)} The branching ratio of\, 
$\tilde{b}_a\rightarrow\chi_1^-\,t$\, as a function of $\tan\beta$ for the
various decays $a=1,2$ with $m_{\tilde{b}_1}<m_{\tilde{b}_2}$;
{\bf (b)} As in (a), but as a function of $m_{\tilde{b}_1}$. 
The marked parts of the abscissa in both figures are excluded
by the condition (\ref{eq:necessary}). The fixed
parameters for (a) and (b) are given in the frame.

\item{\bf Fig.3} {\bf (a)} The SUSY-EW corrections
(\ref{eq:correction}) as a function of $m_{\tilde{t}_1}$;
{\bf (b)} As in (a), but as a function of $m_{\tilde{b}_1}$. 
Rest of inputs as in Fig.2.

\item{\bf Fig.4} {\bf (a)} Evolution of the SUSY-EW corrections as a
function of the sbottom mixing angle, ${\theta}_{\tilde{b}}$, within 
its allowed range;
{\bf (b)} As in (a), but  as a function of the
stop mixing angle, ${\theta}_{\tilde{t}}$. 
Remaining inputs are as in Fig.2
    
\item{\bf Fig.5} {\bf (a)} The SUSY-EW correction as a function of 
$\tan\beta$; {\bf (b)} As in (a), but as a function of $\mu$.
Rest of inputs and notation as in Fig.2.
 
\end{itemize}

\begin{tabular}{c}
\epsfig{file=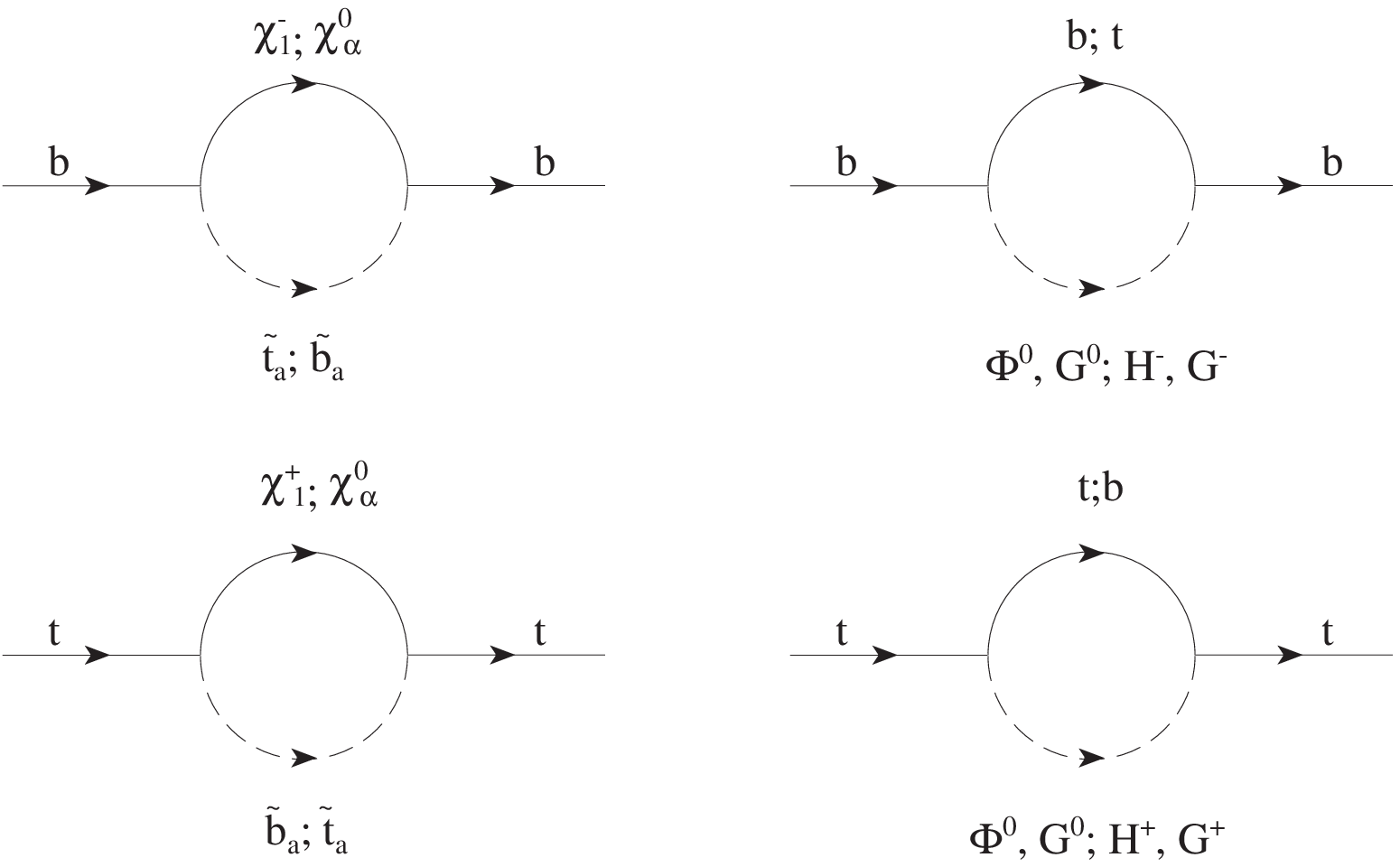,width=10cm}\\
\epsfig{file=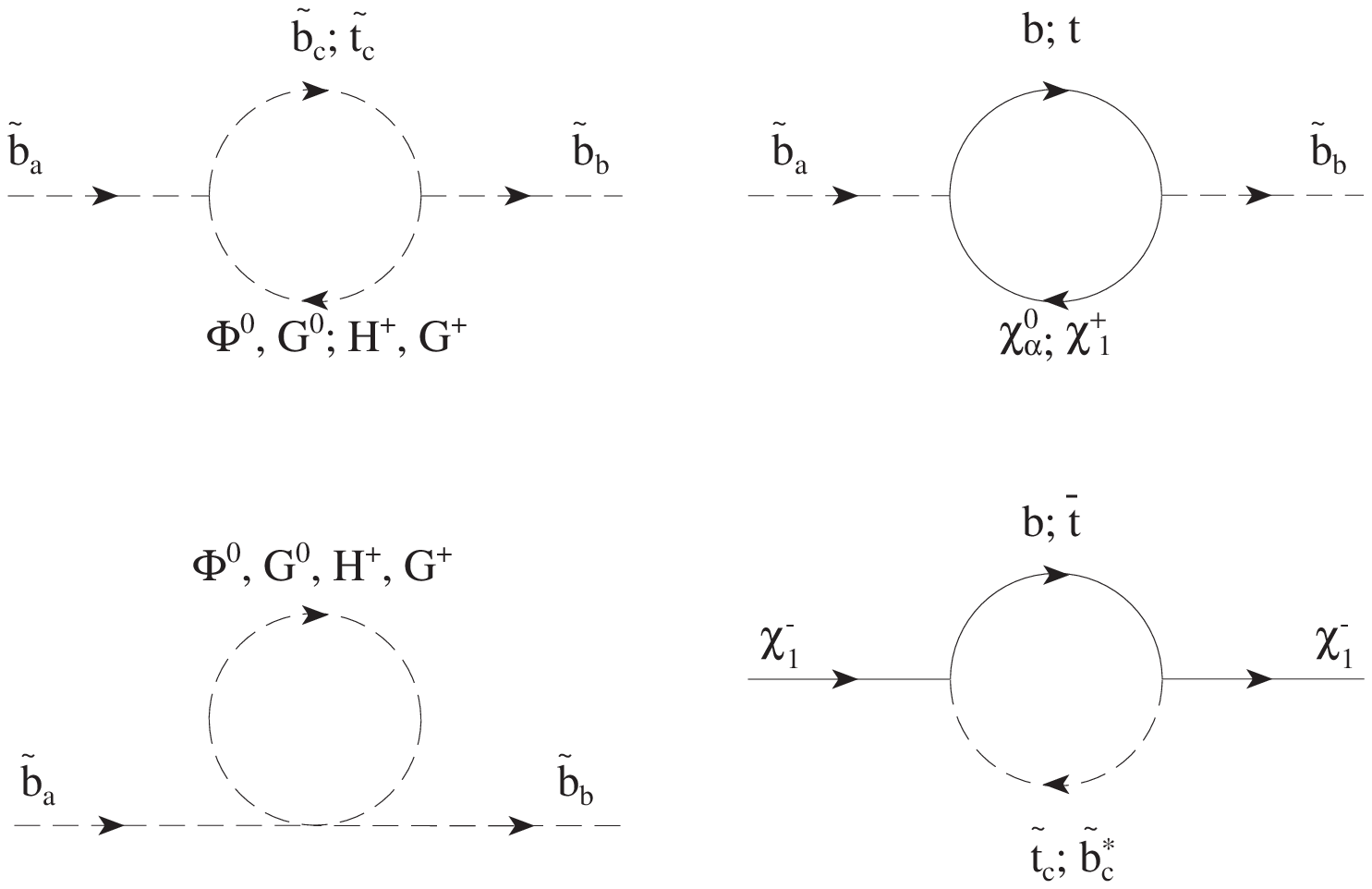,width=10cm}\\
~\\{\Large Fig. 1 (a)}\\~\\~\\
\epsfig{file=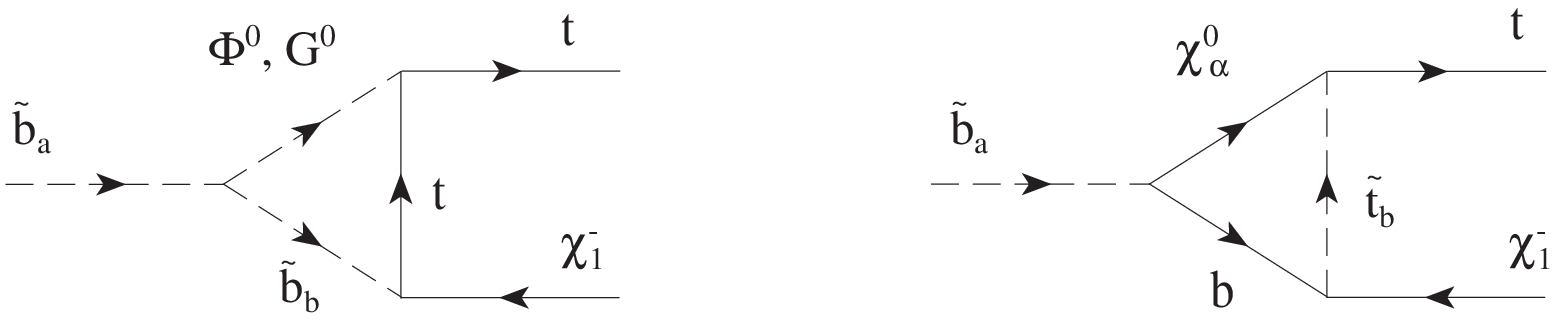,width=10cm}\\
~\\{\Large Fig. 1 (b)}
\end{tabular}

\begin{tabular}{c}
\epsfig{file=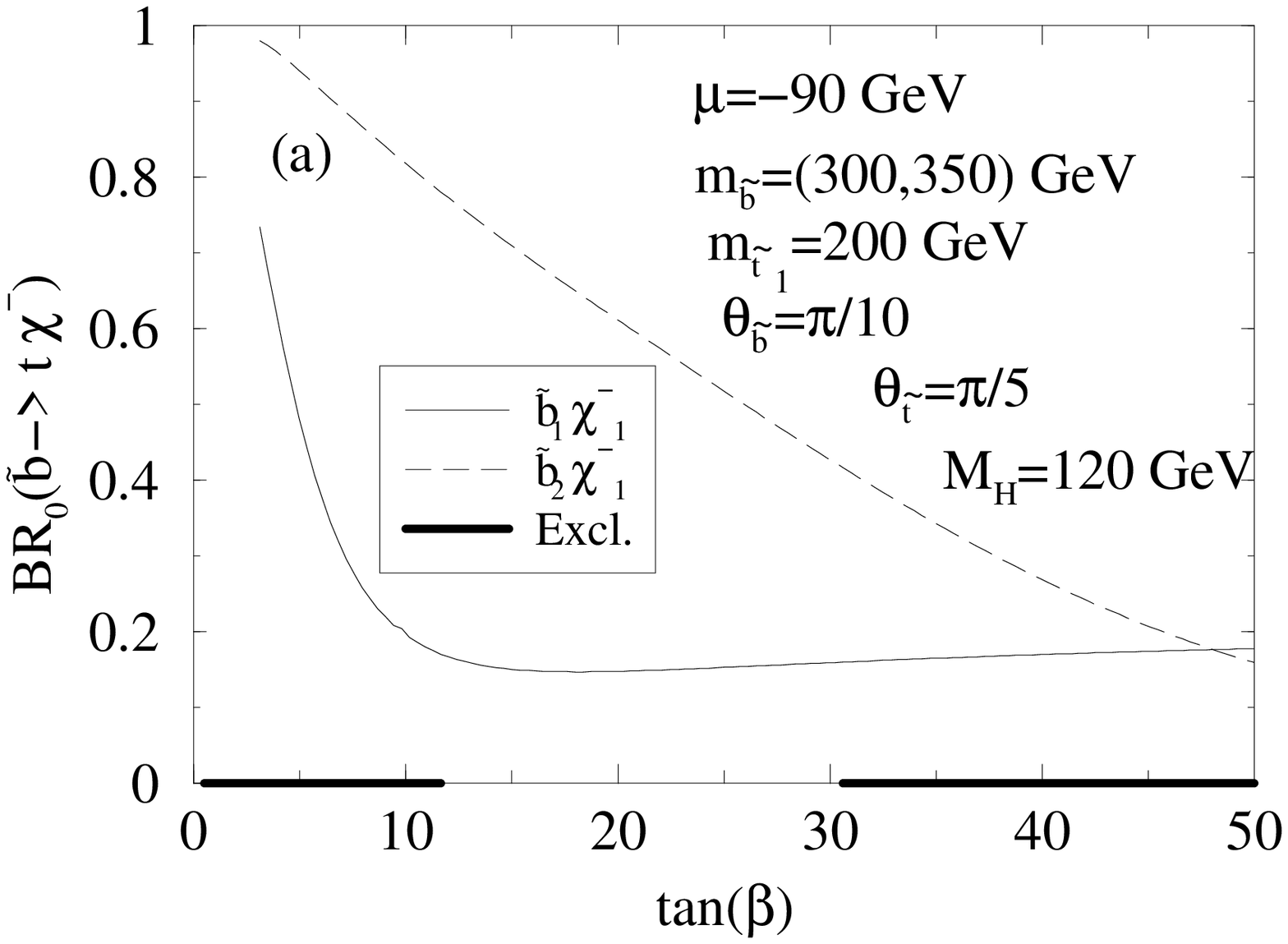,width=10cm}\\
\epsfig{file=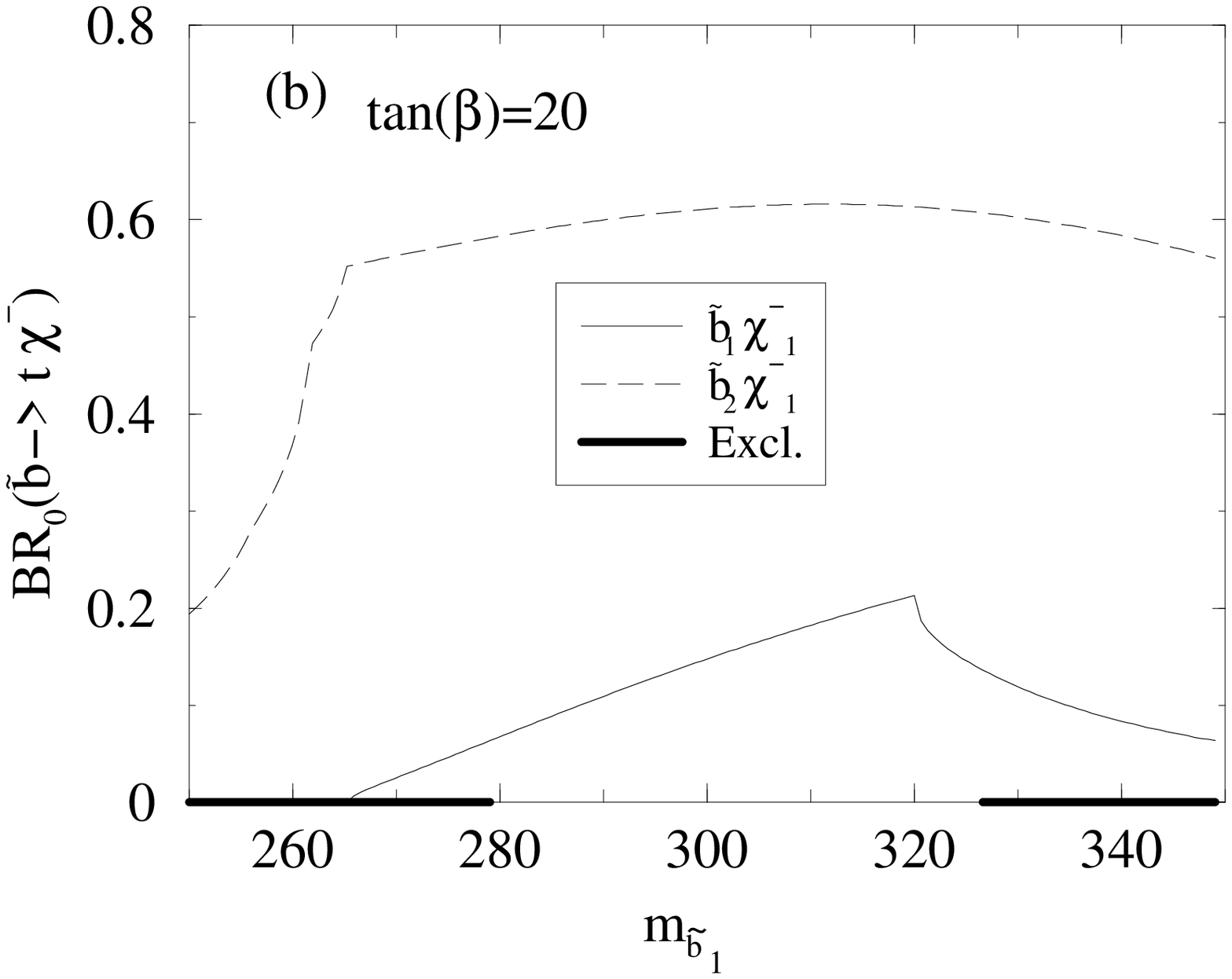,width=10cm}\\
{\Large Fig. 2}
\end{tabular}

\begin{tabular}{c}
\epsfig{file=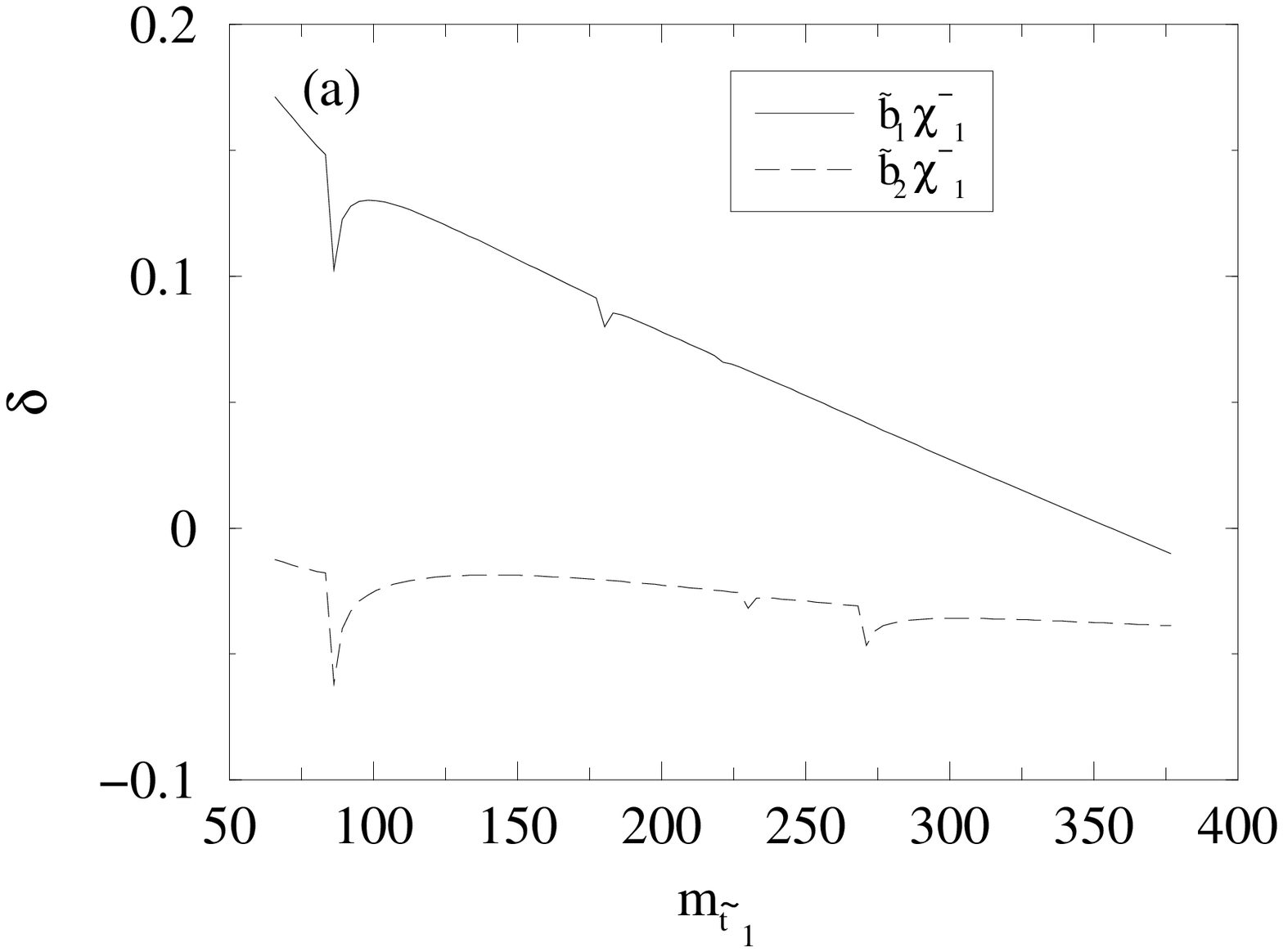,width=10cm}\\
\epsfig{file=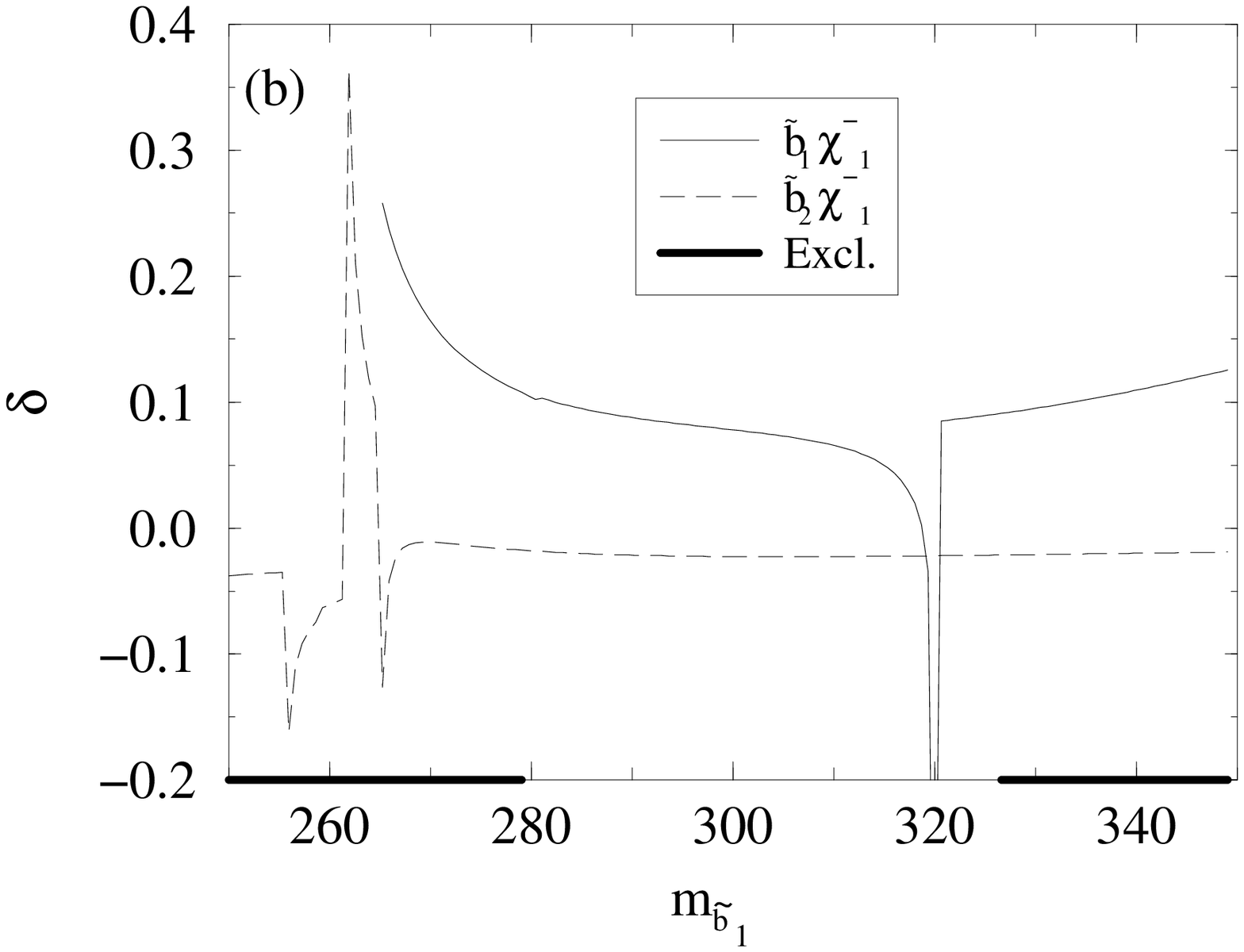,width=10cm}\\
{\Large Fig. 3}
\end{tabular}

\begin{tabular}{c}
\epsfig{file=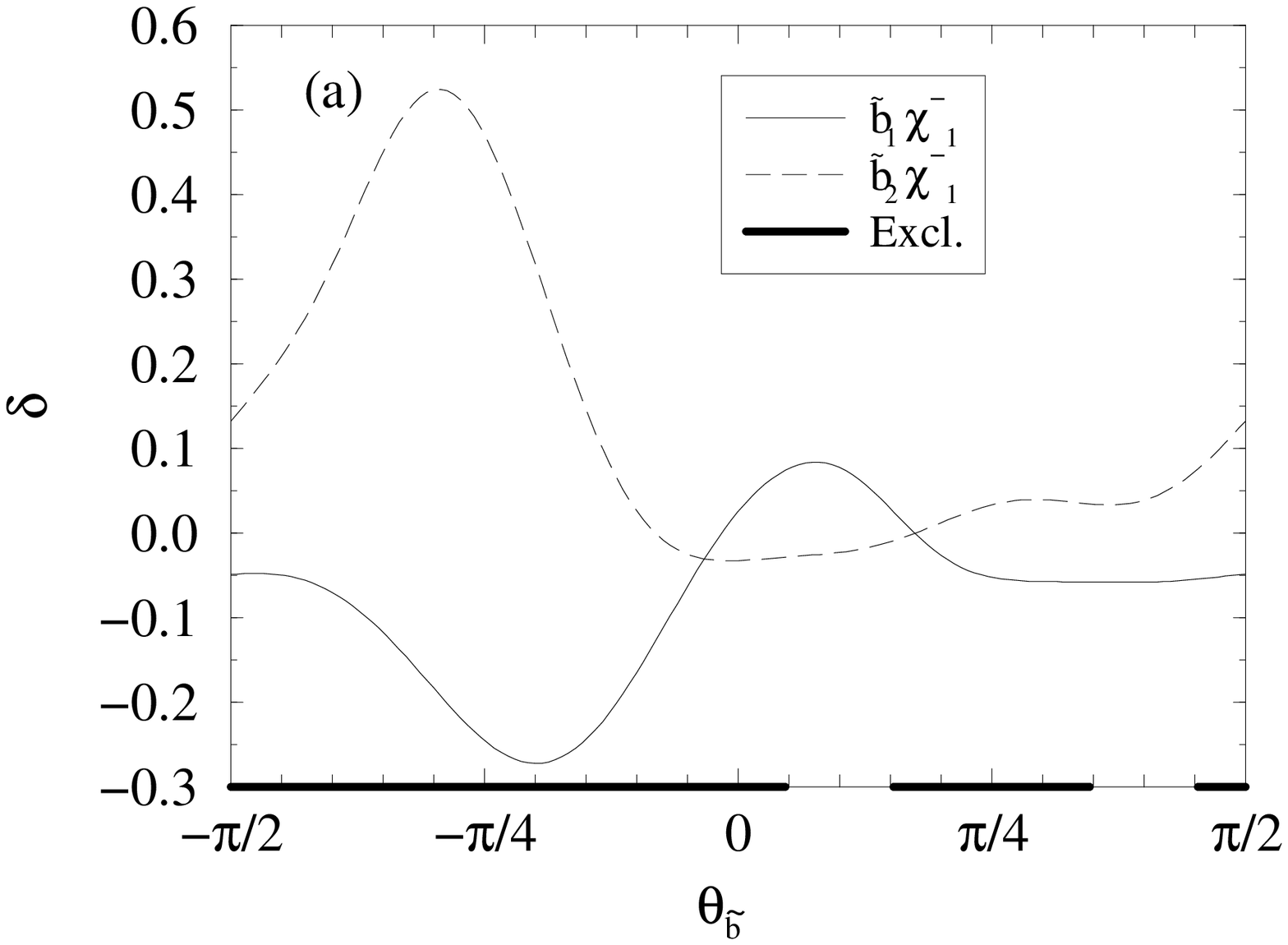,width=10cm}\\
\epsfig{file=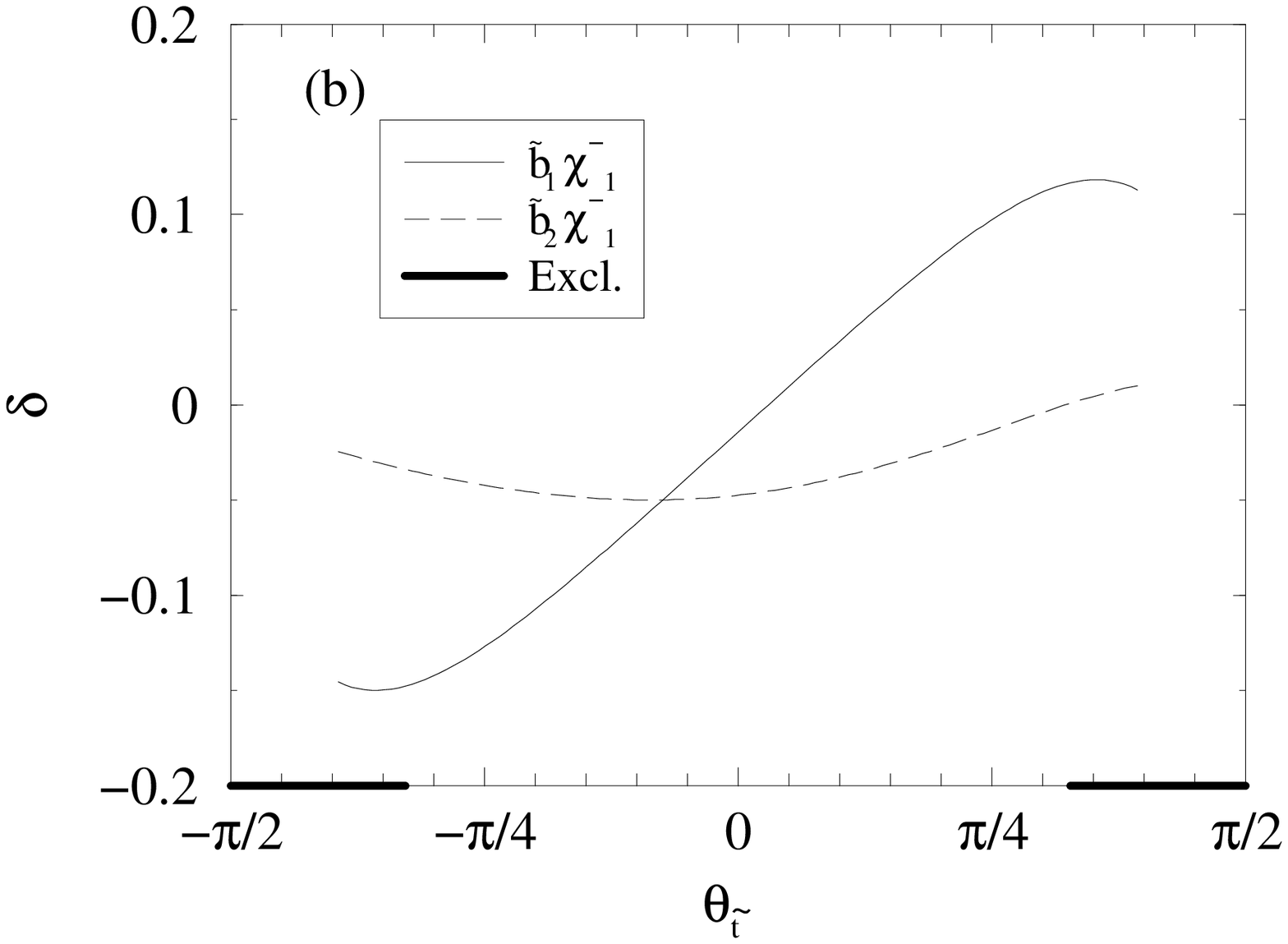,width=10cm}\\
{\Large Fig. 4}
\end{tabular}

\begin{tabular}{c}
\epsfig{file=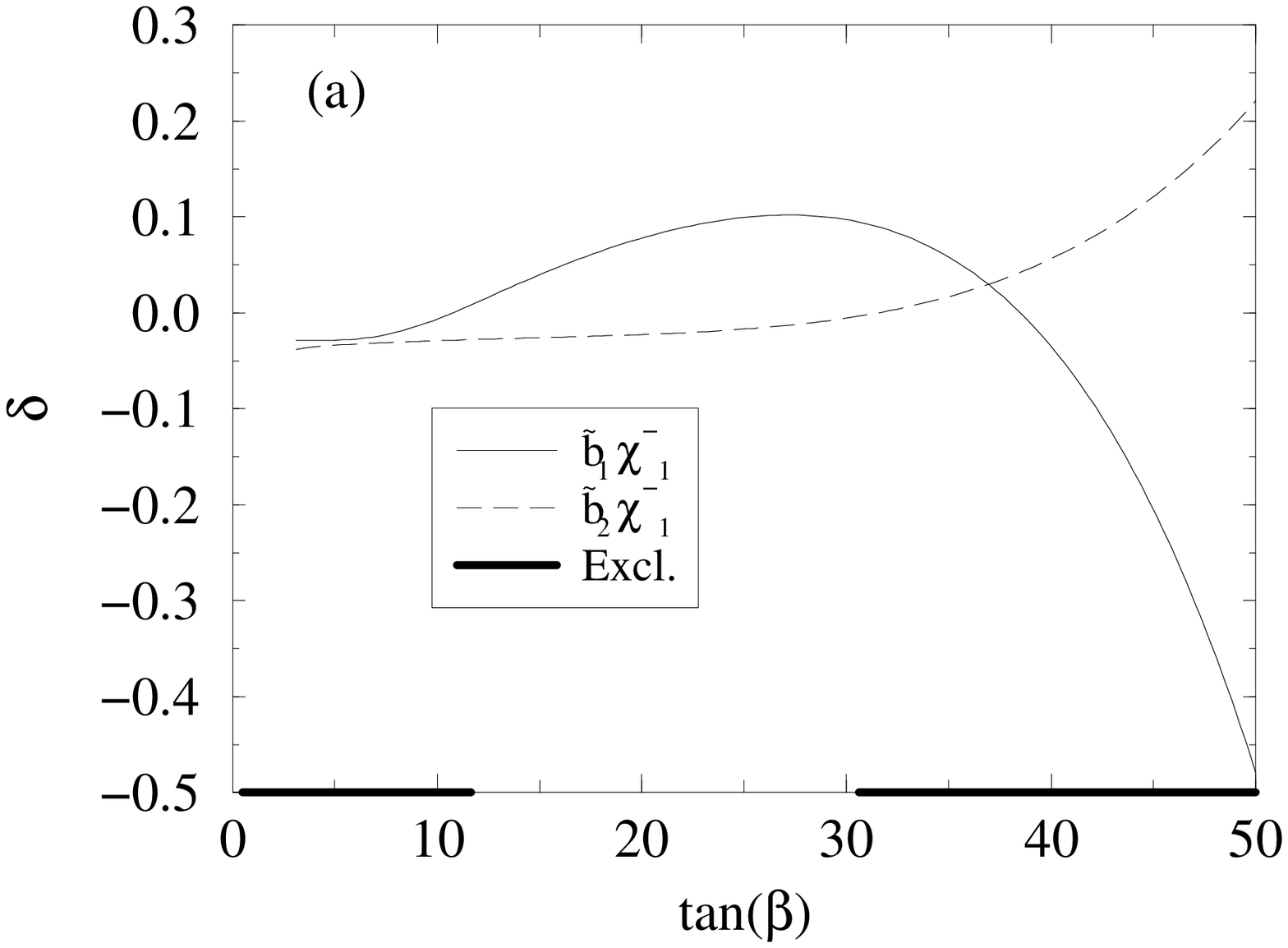,width=10cm}\\
\epsfig{file=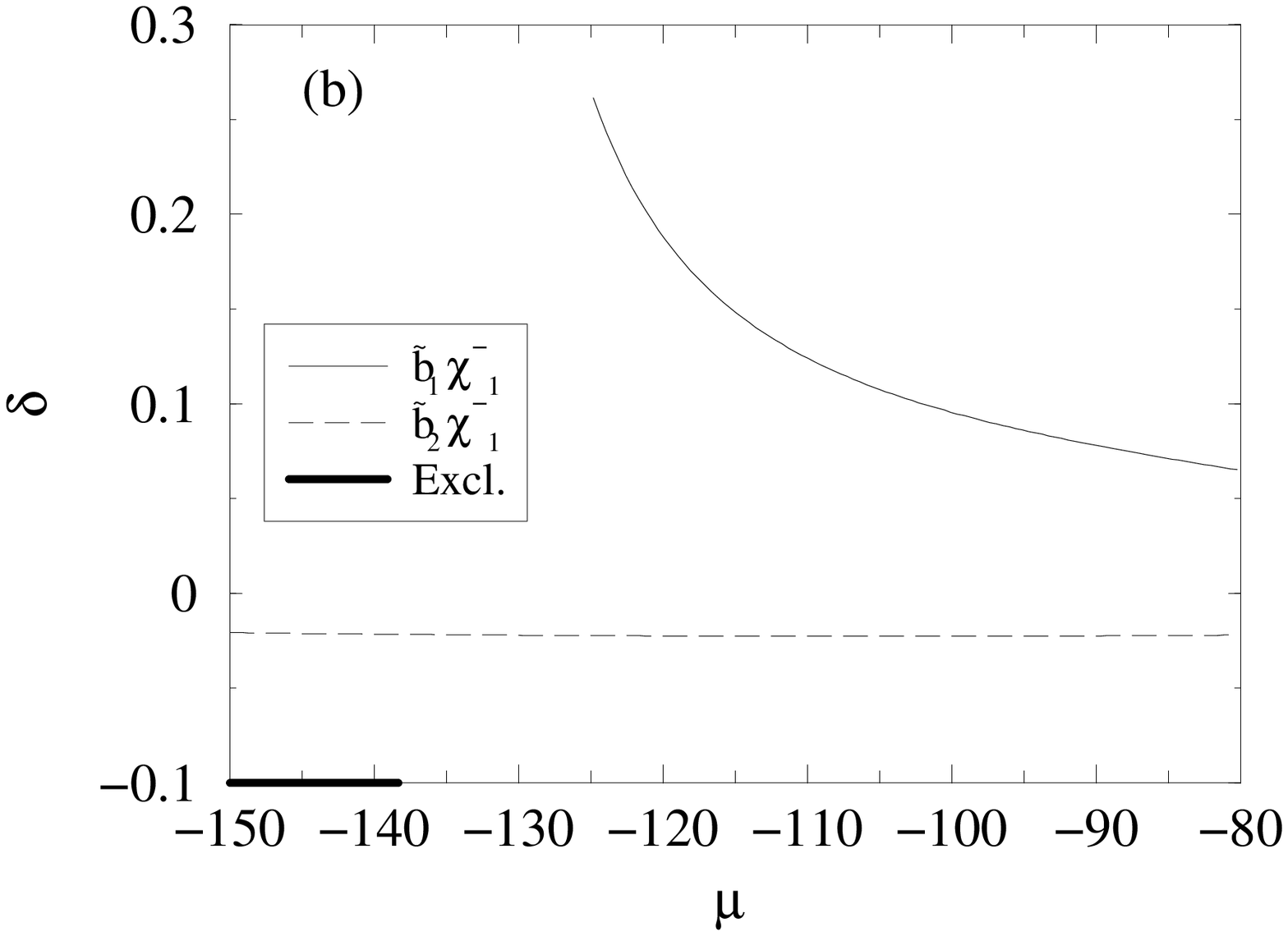,width=10cm}\\
{\Large Fig. 5}
\end{tabular}

\end{document}